\documentclass[twocolumn,amsmath,amssymb,nofootinbib]{revtex4}
\usepackage{graphicx}
\usepackage{eepic}
\usepackage{epic}

\newcommand{\beq}{\begin{equation}}
\newcommand{\eeq}{\end{equation}}
\def\bea{\begin{eqnarray}}
\def\eea{\end{eqnarray}}
\def\nn{\nonumber}

\begin{document} 

\preprint{}

\title{ Large electroweak penguin 
                         contribution in $B\rightarrow K\pi$ modes 
        \footnote{Presented at 2nd International Conference on Flavor
        Physics (ICFP2003),KIAS, Seoul, Oct. 6-11, 2003. This talk
        is based on a work of ref. \cite{YOSHI2}.}}
\author{Tadashi Yoshikawa
         \footnote{e-mail: tadashi@eken.phys.nagoya-u.ac.jp} }
\affiliation{
            Theory Group, KEK, Tsukuba, 305-0801, Japan. \\ 
                         and \\
            Department of Physics, Nagoya University \\
            Nagoya 464-8602, Japan. }

\begin{abstract}
I discuss about a possibility of large electroweak penguin contribution in 
$B\rightarrow K \pi$.  To satisfy several relations
among the branching ratios, we need large electroweak penguin
contribution.  The magnitude is larger than 
the theoretical estimation in the SM so that it may be including 
some new physics effects.      
\end{abstract}

\maketitle

$B\rightarrow K\pi $ modes has been measured\cite{TOMURA} and they will 
be useful informations to understand the CP violation through the KM\cite{KM}
phases and check the standard model(SM).  
If we can directory solve about these modes, it is very elegant way to 
determine the parameters and the weak phase within (or without) the SM. 
However we can not do so because there are 
too many parameters in $B\rightarrow K\pi $ modes. 
So we need understand these modes step by step.
To understand this modes, there are several 
approaches by diagram decomposition\cite{GHLR1}-\cite{FM}, 
QCD factorization\cite{BBNS} and  pQCD\cite{YLS} and so on. 
The largest contribution comes from the gluon penguin. 
If we can deal the contributions from the other diagrams 
as the small parameters, then, there are several relations among 
the averaged branching ratios of $B\rightarrow K\pi$ modes. For example, 
$Br(K^+ \pi^- )/2 Br(K^0 \pi^0 ) \approx 2 Br(K^+ \pi^0)/Br(K^0
\pi^+)$ et al. To examin the relations is very useful to understand
the B decays in the SM\cite{Buras-Fleischer1}. 
However, the recent experiment does not seem to satisfy 
them so well\cite{Buras-Fleischer2,FM,BBNS}. 
When we reconsider these modes to compare with the data, 
we find that the role of a color favored electroweak penguin 
may be important to explain the discrepancy between the relations 
and the experimental data\cite{YOSHI2,GROROS,BFRS,PFl,LICH}. 
So we need to know the informations about the electroweak penguin 
contributions in $B \rightarrow K \pi$ decay modes to understand and
check the SM.   
The ratio between 
gluon and electroweak penguins is estimated in several
works\cite{BBNS, Buras-Fleischer2} and 
it is about $0.14$ as the central value 
but the experimental data suggest that the magnitude should be 
larger than the estimation.  If there is quite large deviation 
in the contribution 
from the electroweak penguin, it may be an evidence of new physics.  

In this talk, I consider how large contribution 
from the electroweak penguins we need in $B \rightarrow K\pi$ modes
and whether it is including some new physics effects.    

Using diagram decomposition and redefinition of the parameters\cite{YOSHI2}, 
the decay amplitudes $A^{XY}$ for $B \rightarrow K^X\pi^Y$ decays 
are written as follows:
\bea
A^{0+} &=& - P |V_{tb}^*V_{ts}|
         \left[ 1 - r_A e^{i\delta^A}e^{i\phi_3}  
                                        \right]  , \\
\sqrt{2} A^{00} &=& 
     - P |V_{tb}^*V_{ts}|
         \left[ 1 - r_{EW}e^{i\delta^{EW}}  
             + r_C e^{i\delta^C}e^{i\phi_3}
                                        \right]  , \nn \\ \\
A^{+-} &=&
            P |V_{tb}^*V_{ts}| 
            \left[ 1 + r_{EW}^C  e^{i\delta^{EWC}} 
             - r_T e^{i\delta^T}e^{i\phi_3}
                                        \right]  , \nn \\ \\
\sqrt{2}A^{+0} &=&  
            P |V_{tb}^*V_{ts}|
                      \left[ 1 + r_{EW} e^{i\delta^{EW}} 
                               + r_{EW}^C e^{i\delta^{EWC}} 
                      \right. \nn \\
       & & \left. 
        - ( r_Te^{i\delta^T} + r_C e^{i\delta^C} + r_A e^{i\delta^A} )
                       e^{i\phi_3} 
                                        \right], 
\eea 
where
$\phi_3$ is the weak phase of $V_{ub}^*V_{us}$, $\delta^X$ is 
the strong phase difference between each diagram and gluon penguin and 
\bea 
r_T &=& \frac{ |T V_{ub}^*V_{us}| }{ |P V_{tb}^*V_{ts}|}, ~~~~
r_C = \frac{ |C V_{ub}^*V_{us}| }{ |P V_{tb}^*V_{ts}|}, \nn \\ 
r_{EW} &=& \frac{ |P_{EW}|}{ |P|}, ~~~~ 
r_{EW}^C = \frac{ |P_{EW}^C|}{ |P|}, \nn \\
r_A &=& \frac{ |A V_{ub}^*V_{us}| }{ |P V_{tb}^*V_{ts}|}. \nn 
\eea
$T$ is a color favored tree amplitude, $C$ is a Color suppressed 
tree, $A$ is an annihilation, $P$  
is a gluonic penguin, $P_{EW}$ is a color 
favored electroweak penguin, and $P_{EW}^C$ is a color suppressed 
electroweak penguin.  
We assume as the hierarchy of the ratios that 
$ 1 > r_T, r_{EW} > r_C, r_{EW}^C > r_A $ ~\cite{GHLR}. 
$|P/T|$ was estimated about $0.1$ in \cite{YOSHI}
\footnote{ Note that this ratio $|P/T|$ does not include CKM factors. } 
by considering the 
$B\rightarrow \pi \pi$ and it was also shown by the ratio of branchings 
of $B^+ \rightarrow \pi^0 \pi^+$ and $B^+ \rightarrow K^0 \pi^+$
\cite{LU,CONV1,CONV2}.  In $B\rightarrow K\pi$ mode, the tree type diagram 
is suppressed by KM factor $V_{ub}^*V_{us}$ and 
$r_T \sim |T/P| \times \lambda^2 R_b \sim 0.2 $, where cabbibo angle 
$\lambda = 0.22$ and we used $R_b = \sqrt{\rho^2+\eta^2} \sim 0.4 $. 
$r_C$ and $r_{EW}^C$ are suppressed by color factor from $r_T$ and $r_{EW}$. 
Comparing the Wilson coefficients which correspond to the diagrams 
under the factorization method, 
we assume that  $r_C \sim 0.1 r_{T}$ and $r_{EW}^C \sim 0.1 r_{EW}$
\cite{LU,BBNS}.  We can also assume that the magnitude of $r_{EW}$ is
$O(0.1)$ by considering the estimation by factorization method.  
$r_A$ could be negligible because it should have B meson 
decay constant and it works as a suppression factor $f_B/M_B$.  
So in this talk, I assume that $r_T, r_{EW} \sim O(0.1)$ and $r_C,
r_{EW}^C \sim O(0.01) $ and 
according to this assumption, 
we will neglect the $r^2$ 
terms including $r_C, r_A$ and $r_{EW}^C$.     
Then, the averaged branching ratios are 
\bea
\bar{B}^{0+}
       &=& |P|^2 |V_{tb}^*V_{ts}|^2 
             \left[ 1 - 2 r_A \cos\delta^A\cos\phi_3  \right],
\label{B0+} \\[5mm]
2 \bar{B}^{00}
      &=& |P|^2 |V_{tb}^*V_{ts}|^2 
                \left[ 1 + r_{EW}^2 
                       - 2 r_{EW} \cos\delta^{EW} \right. \nn \\ 
      & & ~~~~~~ \left.+ 2 r_C \cos\delta^C\cos\phi_3
                \right], \\[5mm]
\bar{B}^{+-}
       &=& |P|^2 |V_{tb}^*V_{ts}|^2 
                \left[ 1 + r_T^2 - 2 r_T \cos\delta^T\cos\phi_3 
                \right. \nn \\
       & & ~~~~~~ \left. + 2 r_{EW}^C \cos\delta^{EWC} 
                \right], \\[5mm]
2 \bar{B}^{+0}
       &=& |P|^2 |V_{tb}^*V_{ts}|^2 
                \left[ 1 +  r_{EW}^2 + r_T^2 
                       + 2 r_{EW} \cos\delta^{EW} \right. \nn \\ 
       & & ~~+ 2 r_{EW}^C \cos\delta^{EWC}  \nn \\ 
       & & - ( 2 r_T \cos\delta^T + 2 r_C \cos\delta^C 
                         + 2 r_A \cos\delta^A )\cos\phi_3 \nn \\
       & & ~~  \left.
                - 2 r_{EW} r_T \cos(\delta^{EW}-\delta^{T})\cos\phi_3
                \right]. 
\label{B+0}
\eea

\begin{table*}[tbhp]
\begin{center}
\begin{tabular}{|c|c|c|c|c|}\hline
  & CLEO\cite{KPCLEO}
  & Belle\cite{TOMURA,KPBELLE} & BaBar\cite{KPBABAR1,KPBABAR2} & Average \\
\hline
$Br(B^+ \rightarrow K^0 \pi^+) \times 10^{6} $ 
             & 18.8 ${}^{+3.7+2.1}_{-3.3-1.8}$
                          & 22.0 $\pm$ 1.9
                          $\pm$ 1.1
                          & 17.5 ${}^{+1.8}_{-1.7}$ $\pm$ 1.3
                          & 19.6 $\pm$ 1.4  \\[2mm]
$Br(B^0 \rightarrow K^0 \pi^0) \times 10^{6} $ 
             & 12.8 ${}^{+4.0+1.7}_{-3.3-1.4}$
                          & 12.6 $\pm$ 2.4
                          $\pm$ 1.4
                          & 10.4 $\pm$1.5 $\pm$ 0.8
                          & 11.2 $\pm$ 1.4  \\[2mm]
$Br(B^0 \rightarrow K^+ \pi^-) \times 10^{6} $ 
             & 18.0${}^{+2.3+1.2}_{-2.1-0.9}$ 
                          & 18.5 $\pm$1.0
                          $\pm$ 0.7
                          & 17.9 $\pm$ 0.9 $\pm$ 0.7
                          & 18.2 $\pm$ 0.8  \\[2mm]
$Br(B^+ \rightarrow K^+ \pi^0) \times 10^{6} $ 
             & 12.9${}^{+2.4+1.2}_{-2.2-1.1}$ 
                          & 12.8 $\pm$1.4
                          ${}^{+1.4}_{-1.0}$ 
                          & 12.8 ${}^{+1.2}_{-1.1}$ $\pm$ 1.0
                          & 12.8 $\pm$ 1.1  \\
\hline
\end{tabular}
\caption{The experimental data and the average. }
\end{center}
\end{table*}

One can take several ratios between the branching ratios. If all modes are 
gluon penguin dominant, the ratios should be close to $1$. 
The shift from $1$ will depend on the magnitude of $r$s.   
From the averaged values of the recent experimental data in Table 1,  
\bea
\frac{\bar{B}^{+-}}{2\bar{B}^{00}} &=& 0.81 \pm 0.11, 
\label{data+-}\\
\frac{2 \bar{B}^{+0}}{\bar{B}^{0+}} &=& 1.31 \pm 0.15, 
  \\[5mm]
\frac{\tau^+}{\tau^0}\frac{\bar{B}^{+-}}{\bar{B}^{0+}} &=& 1.01 \pm 0.09, \\
\frac{\tau^0}{\tau^+}\frac{\bar{B}^{+0}}{\bar{B}^{00}} &=& 1.05 \pm 0.16, 
                                                     \\[5mm]
\frac{\tau^+}{\tau^0}\frac{2 \bar{B}^{00}}{\bar{B}^{0+}} &=& 1.24 \pm 0.18, \\
\frac{\tau^0}{\tau^+}\frac{2 \bar{B}^{+0}}{\bar{B}^{+-}} &=& 1.30 \pm 0.13,
\label{data+0}  
\eea 
where $\frac{\tau^+}{\tau^0}$ is a lifetime ratio 
between the charged and the neutral $B$ mesons and 
$\tau(B^\pm)/\tau(B^0) = 1.083 \pm 0.017$\cite{PDG}. 
As an example. we can find two ratios 
from eq.(\ref{B0+})-(\ref{B+0}) under the assumption that 
all $r$ is smaller than $1$ and the $r^2$ terms including $r_C, r_A$ 
and $r_{EW}^C$ are neglected,         
\bea
\frac{\bar{B}^{+-}}{2\bar{B}^{00}} &=& \left\{
  1 + 2 r_{EW} \cos\delta^{EW} + 2 r_{EW}^C \cos\delta^{EWC}
               \right. \nn \\ 
  & & \left. - 2 ( r_{T} \cos\delta^{T} + r_C \cos\delta^C )\cos\phi_3  
    + r_{T}^2 \right\}\nn \\
  & & ~~~ -r_{EW}^2 + 4 r_{EW}^2 \cos^2\delta^{EW}, 
\label{B+-B00}\\[4mm]
\frac{2 \bar{B}^{+0}}{\bar{B}^{0+}} &=&
 \left\{ 1 + 2 r_{EW} \cos\delta^{EW} + 2 r_{EW}^C \cos\delta^{EWC} 
    \right. \nn \\
  & & \left. - 2 ( r_{T} \cos\delta^{T} + r_C \cos\delta^C )\cos\phi_3  
  + r_{T}^2 \right\}   \\
  & & ~~~ + r_{EW}^2 
            - 2 r_{EW} r_T \cos(\delta^{EW}-\delta^T)\cos\phi_3, \nn  
\label{B+0B0+}
\eea
The equations are same up to $r_T^2$ term 
and the difference seems to come from $r_{EW}^2$ term. The difference
of experimental data from 1 seems to depend on the sign of $r_{EW}^2$ term.

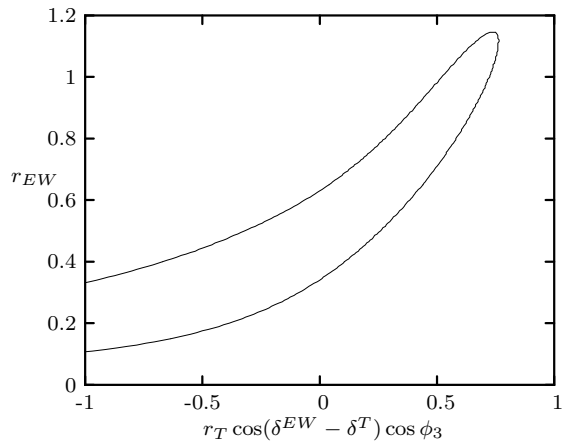
\begin{figure}[htbp]
\begin{center}
\setlength{\unitlength}{0.080450pt}
\begin{picture}(2699,2069)(0,0)
\footnotesize
\thicklines \path(370,249)(411,249)
\thicklines \path(2576,249)(2535,249)
\put(329,249){\makebox(0,0)[r]{ 0}}
\thicklines \path(370,539)(411,539)
\thicklines \path(2576,539)(2535,539)
\put(329,539){\makebox(0,0)[r]{ 0.2}}
\thicklines \path(370,829)(411,829)
\thicklines \path(2576,829)(2535,829)
\put(329,829){\makebox(0,0)[r]{ 0.4}}
\thicklines \path(370,1119)(411,1119)
\thicklines \path(2576,1119)(2535,1119)
\put(329,1119){\makebox(0,0)[r]{ 0.6}}
\thicklines \path(370,1408)(411,1408)
\thicklines \path(2576,1408)(2535,1408)
\put(329,1408){\makebox(0,0)[r]{ 0.8}}
\thicklines \path(370,1698)(411,1698)
\thicklines \path(2576,1698)(2535,1698)
\put(329,1698){\makebox(0,0)[r]{ 1}}
\thicklines \path(370,1988)(411,1988)
\thicklines \path(2576,1988)(2535,1988)
\put(329,1988){\makebox(0,0)[r]{ 1.2}}
\thicklines \path(370,249)(370,290)
\thicklines \path(370,1988)(370,1947)
\put(370,166){\makebox(0,0){-1}}
\thicklines \path(922,249)(922,290)
\thicklines \path(922,1988)(922,1947)
\put(922,166){\makebox(0,0){-0.5}}
\thicklines \path(1473,249)(1473,290)
\thicklines \path(1473,1988)(1473,1947)
\put(1473,166){\makebox(0,0){ 0}}
\thicklines \path(2025,249)(2025,290)
\thicklines \path(2025,1988)(2025,1947)
\put(2025,166){\makebox(0,0){ 0.5}}
\thicklines \path(2576,249)(2576,290)
\thicklines \path(2576,1988)(2576,1947)
\put(2576,166){\makebox(0,0){ 1}}
\thicklines \path(370,249)(2576,249)(2576,1988)(370,1988)(370,249)
\put(22,1218){\makebox(0,0)[l]{$r_{EW}$}}
\put(1473,42){\makebox(0,0){ $ 
         r_{T} \cos(\delta^{EW}-\delta^{T})\cos \phi_3 $}}
\thinlines \path(2300,1908)(2300,1908)(2306,1901)(2311,1894)
(2311,1887)(2311,1879)(2317,1872)(2317,1865)(2311,1858)(2311,1850)
(2311,1843)(2311,1836)(2311,1829)(2311,1821)(2306,1814)(2306,1807)
(2306,1800)(2300,1792)(2300,1785)(2300,1778)(2295,1771)(2295,1763)
(2289,1756)(2289,1749)(2289,1742)(2284,1734)(2284,1727)(2278,1720)
(2278,1713)(2273,1705)(2273,1698)(2267,1691)(2262,1684)(2262,1676)
(2256,1669)(2256,1662)(2251,1655)(2251,1647)(2245,1640)(2240,1633)
(2240,1626)(2234,1618)(2229,1611)(2229,1604)(2223,1597)(2218,1589)
(2218,1582)(2212,1575)(2206,1568)(2206,1560)(2201,1553)
\thinlines \path(2201,1553)(2195,1546)(2195,1539)(2190,1532)(2184,1524)
(2179,1517)(2179,1510)(2173,1503)(2168,1495)(2162,1488)(2162,1481)
(2157,1474)(2151,1466)(2146,1459)(2140,1452)(2140,1445)(2135,1437)
(2129,1430)(2124,1423)(2118,1416)(2113,1408)(2113,1401)(2107,1394)
(2102,1387)(2096,1379)(2091,1372)(2085,1365)(2080,1358)(2074,1350)
(2069,1343)(2069,1336)(2063,1329)(2058,1321)(2052,1314)(2047,1307)
(2041,1300)(2036,1292)(2030,1285)(2025,1278)(2019,1271)(2013,1263)
(2008,1256)(2002,1249)(1997,1242)(1991,1234)(1986,1227)(1980,1220)
(1975,1213)(1969,1205)(1964,1198)(1958,1191)
\thinlines \path(1958,1191)(1953,1184)(1947,1176)(1936,1169)(1931,1162)
(1925,1155)(1920,1147)(1914,1140)(1909,1133)(1903,1126)(1898,1119)
(1887,1111)(1881,1104)(1876,1097)(1870,1090)(1865,1082)(1859,1075)
(1848,1068)(1843,1061)(1837,1053)(1831,1046)(1820,1039)(1815,1032)
(1809,1024)(1804,1017)(1793,1010)(1787,1003)(1782,995)(1771,988)
(1765,981)(1760,974)(1749,966)(1743,959)(1732,952)(1727,945)(1721,937)
(1710,930)(1705,923)(1694,916)(1688,908)(1677,901)(1672,894)(1661,887)
(1655,879)(1644,872)(1638,865)(1627,858)(1616,850)(1611,843)(1600,836)
(1589,829)
\thinlines \path(1589,829)(1583,821)(1572,814)(1561,807)(1556,800)
(1545,792)(1534,785)(1523,778)(1512,771)(1501,763)(1490,756)(1484,749)
(1473,742)(1462,734)(1445,727)(1434,720)(1423,713)(1412,705)(1401,698)
(1390,691)(1374,684)(1363,677)(1352,669)(1335,662)(1324,655)(1308,648)
(1297,640)(1280,633)(1263,626)(1247,619)(1236,611)(1219,604)(1203,597)
(1181,590)(1164,582)(1148,575)(1126,568)(1109,561)(1087,553)(1065,546)
(1048,539)(1021,532)(999,524)(977,517)(949,510)(922,503)(894,495)(866,488)
(839,481)(806,474)(773,466)
\thinlines \path(773,466)(734,459)(695,452)(657,445)(613,437)(569,430)
(519,423)(464,416)(403,408)(370,404)
\thinlines \path(2300,1908)(2273,1908)(2273,1908)(2256,1901)
(2245,1894)(2234,1887)
(2223,1879)(2218,1872)(2206,1865)(2201,1858)(2190,1850)(2184,1843)
(2173,1836)(2168,1829)(2162,1821)(2151,1814)(2146,1807)(2140,1800)
(2135,1792)(2124,1785)(2118,1778)(2113,1771)(2107,1763)(2102,1756)
(2091,1749)(2085,1742)(2080,1734)(2074,1727)(2069,1720)(2063,1713)
(2052,1705)(2047,1698)(2041,1691)(2036,1684)(2030,1676)(2025,1669)
(2013,1662)(2008,1655)(2002,1647)(1997,1640)(1991,1633)(1986,1626)
(1975,1618)(1969,1611)(1964,1604)(1958,1597)(1953,1589)(1947,1582)
(1936,1575)(1931,1568)(1925,1560)(1920,1553)
\thinlines \path(1920,1553)(1914,1546)(1909,1539)(1898,1532)(1892,1524)
(1887,1517)(1881,1510)(1870,1503)(1865,1495)(1859,1488)(1854,1481)
(1843,1474)(1837,1466)(1831,1459)(1826,1452)(1815,1445)(1809,1437)
(1804,1430)(1793,1423)(1787,1416)(1782,1408)(1771,1401)(1765,1394)
(1754,1387)(1749,1379)(1743,1372)(1732,1365)(1727,1358)(1716,1350)
(1710,1343)(1699,1336)(1694,1329)(1683,1321)(1677,1314)(1666,1307)
(1661,1300)(1649,1292)(1638,1285)(1633,1278)(1622,1271)(1611,1263)
(1605,1256)(1594,1249)(1583,1242)(1578,1234)(1567,1227)(1556,1220)
(1545,1213)(1534,1205)(1528,1198)(1517,1191)
\thinlines \path(1517,1191)(1506,1184)(1495,1176)(1484,1169)(1473,1162)
(1462,1155)(1451,1147)(1440,1140)(1429,1133)(1412,1126)(1401,1119)
(1390,1111)(1379,1104)(1368,1097)(1352,1090)(1341,1082)(1330,1075)
(1313,1068)(1302,1061)(1285,1053)(1274,1046)(1258,1039)(1247,1032)
(1230,1024)(1214,1017)(1203,1010)(1186,1003)(1170,995)(1153,988)
(1137,981)(1120,974)(1103,966)(1087,959)(1070,952)(1054,945)(1037,937)
(1021,930)(999,923)(982,916)(966,908)(944,901)(927,894)(905,887)
(888,879)(866,872)(844,865)(822,858)(800,850)(778,843)(756,836)(734,829)
\thinlines \path(734,829)(712,821)(690,814)(662,807)(640,800)(613,792)
(591,785)(563,778)(535,771)(508,763)(480,756)(453,749)(425,742)
(392,734)(370,729)
\end{picture}
\caption{The allowed region at $1\sigma $ level on 
$(r_{EW}, ~r_T \cos(\delta^{EW}~-~\delta^{T}) \cos \phi_3)$ plane. }
    \label{fig:1}
\end{center}
\end{figure}

If we can neglect all $r^2$ terms, then there are a few relations among 
the ratios as following 
\bea
\frac{\bar{B}^{+-}}{2\bar{B}^{00}} &-& \frac{2
\bar{B}^{+0}}{\bar{B}^{0+}} 
                             = 0,
                                   \\[3mm]
\frac{2 \bar{B}^{+0}}{\bar{B}^{0+}}
 - \frac{\tau^+}{\tau^0}\frac{\bar{B}^{+-}}{\bar{B}^{0+}} &+& 
\frac{\tau^+}{\tau^0}\frac{2 \bar{B}^{00}}{\bar{B}^{0+}} -1 = 0 . 
\label{BISO}
\eea 
However, the experimental data listed in eqs.(\ref{data+-})-(\ref{data+0})
do not satisfy these relations so well. According 
to the experimental data, 
$\frac{\bar{B}^{+-}}{2\bar{B}^{00}}$ seems to be smaller than 1 but 
$\frac{2 \bar{B}^{+0}}{\bar{B}^{0+}}$ be larger than 1. So it shows 
there is a discrepancy between them.  
$r_T^2$ term does not seem to contribute to the ratios so strongly. 
The second relation corresponds to the isospin relation 
at the first order of $r$.  
The discrepancy of relation (\ref{BISO}) from $0$ also comes from 
$r_{EW}^2$ term. 
The differences are
\begin{widetext} 
\bea
\frac{2 \bar{B}^{+0}}{\bar{B}^{0+}} - \frac{\bar{B}^{+-}}{2\bar{B}^{00}} = 
   2 r_{EW}^2 - 2 r_{EW} r_{T} \cos(\delta^{EW}-\delta^{T})\cos\phi_3
              - 4 r_{EW}^2 \cos^2\delta^{EW} = 0.50 \pm 0.19 ,
\label{B+0B0+MB+-B00} \\[3mm]
\frac{2 \bar{B}^{+0}}{\bar{B}^{0+}}
 - \frac{\tau^+}{\tau^0}\frac{\bar{B}^{+-}}{\bar{B}^{0+}} + 
\frac{\tau^+}{\tau^0}\frac{2 \bar{B}^{00}}{\bar{B}^{0+}} -1 = 
2 r_{EW}^2 - 2 r_{EW} r_{T} \cos(\delta^{EW}-\delta^{T})\cos\phi_3 
                                                      = 0.54 \pm 0.25 , 
\label{B+0MB+-PB00M1}
\eea 
\end{widetext}
so that one can find the electroweak penguin contributions may be large. 
All terms are including $r_{EW}$ and the deviation of the relation from $0$
is finite. Here the error are determined by adding quadratically all errors. 
Using the other relation as following  
\bea
 & & \frac{\bar{B}^{+-}}{2\bar{B}^{00}}
 - \frac{\tau^0}{\tau^+}\frac{\bar{B}^{+0}}{\bar{B}^{00}} + 
\frac{\tau^+}{\tau^0}\frac{2 \bar{B}^{00}}{\bar{B}^{0+}} - 1 \nn \\
& & = 
- 4 r_{EW} \cos\delta^{EW} 
          + 2 r_{EW} r_{T} \cos(\delta^{EW}-\delta^{T})\cos\phi_3 \nn \\
& & = 0.00 \pm 0.26 , 
\label{B+-MB+0PB00M1}
\eea 
we can solve them about $r_{EW}$ and if we can respect the central values,
the solutions are
\bea
 & & (r_{EW},  ~\cos\delta^{EW}, 
   ~r_{T} \cos(\delta^{EW}-\delta^{T})\cos\phi_3 ) \nn \\
& &  = 
( 0.26,  -0.38, -0.75 ) ~\mbox{and} ~( 0.69, 0.21, 0.41 ) . 
\label{solution}
\eea
This solution show that large electroweak penguin contribution (but 
$r_{T} \cos(\delta^{EW}-\delta^{T})\cos\phi_3$ is too large because 
$r_T$ was estimated around 0.2 by the other methods.)   
The allowed region of $r_{EW}$ 
and $r_{T} \cos(\delta^{EW}-\delta^{T})\cos\phi_3 $ at 
$1 \sigma $ level for eqs.(\ref{B+0B0+MB+-B00})-(\ref{B+-MB+0PB00M1}) 
is shown in Fig.\ref{fig:1}. 

From this result, we find that the smaller $r_{EW}$ will favored 
a larger $|r_T \cos(\delta^{EW}~-~\delta^{T}) \cos \phi_3|$ term 
with negative sign. 
However such large $r_{T}$ is disfavored by the rough estimation of $r_T$ 
which is around 0.2.    
Even if $|r_T \cos(\delta^{EW}~-~\delta^{T}) \cos \phi_3|$ is within 0.2, 
then $r_{EW}$ will also be larger than 0.3 and $r_{EW}$ will be larger than
$r_T$.  This is showing that there is 
a possibility of large electroweak penguin contribution and it may be
an evidence of new physics effects. 
Roughly speaking, the shift of eqs.(\ref{data+-})-(\ref{data+0}) from $1$ 
seem to depend on the $r_{EW}^2$ term and the sign. 
The contributions from tree diagram are not so large except for the cross 
term with the electroweak penguin because 
$\frac{\tau^+}{\tau^0}\frac{\bar{B}^{+-}}{\bar{B}^{0+}}$ is quite near $1$. 
To fix the solution or confirm the large electroweak penguin contribution, 
we need higher accurate data.      

In this talk, I discussed about  
a possibility of large electro-weak penguin contribution in 
$B\rightarrow K \pi$ from recent experimental data. 
The several relations among the branching ratios which realize when 
the contributions from tree type and electroweak penguin are small 
compared with the gluon penguin do not satisfy the data. 
The difference comes from the $r^2$ 
terms and the main contribution comes from electroweak penguin. 
We find that the contribution from electroweak penguin 
may be larger than from tree diagrams to explain the experimental data. 
If the magnitude estimated from experiment is quite large compared 
with the theoretical estimation which is usually smaller 
than tree contributions\cite{NEU,FM,BBNS}, then it may be including 
some new physics effects. In this analysis, we find that who can have 
some contribution from new physics is 
the color favored Z penguin type diagram (ZFCNC\cite{YOSHI2,AH,BFRS}) 
which is the process 
$\pi^0$ goes out from $B-K$ current.

\section*{Acknowledgments}
I would like to thank the organizers and members of KIAS for their
hospitality during the conference and C.S. Kim for inviting me to Yonsei U.

\end{document}